\documentclass{ws-procs9x6}

\newcommand{\vk}{\mbox {\boldmath $k$\unboldmath}}
\newcommand{\vq}{\mbox {\boldmath $q$\unboldmath}}
\def\nn{\nonumber}

\begin{document}

\title{Strangeness production {\it via} electromagnetic probes: 
40 years later\footnote{Invited talk given at the
International Symposium on Electrophoto-production of Strangeness
on Nucleons and Nuclei, Sendai, Japan, 
June 16-18, 2003; World Scientific ({\it to appear}).}}

\author{B. SAGHAI}

\address{D\'epartement d'Astrophysique, de Physique des Particules, de Physique 
Nucl\'eaire\\ 
et de l'Instumentation Associ\'ee, CEA/Saclay, 91191 Gif-sur-Yvette, 
France\\
E-mail: bsaghai@cea.fr}


\maketitle
\abstracts{
A brief review of the associated strangeness electromagnetic production is 
presented.
Very recent $K^+ \Lambda$ photoproduction data on the proton from 
threshold up to $E_\gamma ^{lab}$ = 2.6 GeV are interpreted within 
a chiral constituent quark formalism, which embodies all known nucleonic 
and hyperonic resonances. The preliminary results of this work are reported here.
}
\section{Introduction}
%
This conference witnessed the advent of thousands of data points from 
JLab/CLAS\cite{CLAS} and ELSA/SAPHIR\cite{Saphir} on the $K^{+} \Lambda$
and $K^{+} \Sigma^\circ$ photoproduction on the proton. Moreover, the recent data
with polarized photon beam from Spring-8/LEPS\cite{LEPS} and with electrons  
from JLab\cite{Elec} announce the start of a new era in this field,
ringing the moment of the truth for phenomenologists!

Actually, this journey started some 40 years ago (see e.g. Ref.\cite{AS}), 
with the pionner and significant 
work performed by Thom\cite{Thom}. The data which are now becoming available, 
were anticipated some 20 years ago and gave a new momentum to the phenomenological
investigations\cite{AS,CoAd}. Those studies, based on the Feynman diagrammatic 
technique, included {\it s-},{\it u-}, and {\it t-}channel contributions, 
and produced models differing mainly in their content of baryon
resonances. Later, more sophistications were 
introduced\cite{NC,SL,OS,MB,Yonsei,Janssen}, the most significant ones
being: 

{\it i)} Introduction of spin- 3/2 and 5/2 nucleonic\cite{SL,Yonsei} 
and spin-3/2 hyperonic resonances\cite{OS}. This latter would not have been possible 
without the incorporation\cite{OS} of the so-called {\it off-shell effects} 
inherent to the fermions with spin $\ge 3/2$, which of course also applies
to the relevant nucleon resonances.
 
{\it ii)} Introduction of hadronic form factors at strong vertices and preserving the gauge 
invariance of the amplitudes\cite{SFF}.
 
The well known main difficulty in the kaon production, compared to the $\pi$
and $\eta$ cases, is that the reaction mechanism here is not dominated
by a small number of resonances. This fact implies that we need to embody in the
model contributions from a large number of resonances, the known
ones being shown in Table~I. Given that, according to the spin of the resonances,
one needs 1 to 5 free parameters
per resonance; it is obvious that a meaningful study
of the resonance content of the underlying reaction mechanism is excluded within
such approaches.

\vspace{2mm}
%
%
\noindent {Table I. Baryon resonances\cite{PDG} with mass 
$M_{N^*}\leq$ 2.5 GeV. Notations are $L_{2I~2J}(mass)$ and 
$L_{I~2J}(mass)$ for $N^*$ and $Y^*$, respectively.}
\vspace{-5.3mm}
\begin{table}[h!]
%
%
\begin{tabular}{|c|l|l|}
\hline 
 Baryon &  Three \& four star resonances &  One \& two star resonances \\
\hline
  & $S_{11}(1535)$, $S_{11}(1650)$, & $S_{11}(2090)$, \\
 & $P_{11}(1440)$, $P_{11}(1710)$, $P_{13}(1720)$,
 & $P_{11}(2100)$, $P_{13}(1900)$, \\
{\large $N^*$} & $D_{13}(1520)$, $D_{13}(1700)$, $D_{15}(1675)$, 
 & $D_{13}(2080)$, $D_{15}(2200)$, \\
 & $F_{15}(1680)$,  
 & $F_{15}(2000)$, $F_{17}(1990)$, \\
 & $G_{17}(2190)$, $G_{19}(2250)$,  
 &   \\
 & $H_{19}(2220)$, 
 &   \\ [10pt]
 & $S_{01}(1405)$, $S_{01}(1670)$, $S_{01}(1800)$,  
 &  \\
 &  $P_{01}(1600)$, $P_{01}(1810)$, $P_{03}(1890)$, 
 & \\
{\large $\Lambda^*$} & $D_{03}(1520)$, $D_{03}(1690)$, $D_{05}(1830)$,
 & $D_{03}(2325)$, \\
 & $F_{05}(1820)$, $F_{05}(2110)$,  
 & $F_{07}(2020)$,  \\
 & $G_{07}(2100)$, 
 &   \\
 & $H_{09}(2350)$,
 &   \\ [10pt]
 & $S_{11}(1750)$, 
 & $S_{11}(1620)$, $S_{11}(2000)$, \\
 & $P_{11}(1660)$, $P_{11}(1880)$, $P_{13}(1385)$, 
 & $P_{11}(1770)$, $P_{11}(1880)$,  \\ 
{\large $\Sigma^*$} && $P_{13}(1840)$, $P_{13}(2080)$, \\ 
%
 & $D_{13}(1670)$, $D_{13}(1940)$, $D_{15}(1775)$,
 & $D_{13}(1580)$, \\
 & $F_{15}(1915)$, $F_{17}(2030)$. 
 & $F_{15}(2070)$,   \\
 &  
 & $G_{17}(2100)$.  \\
\hline
\end{tabular}
\end{table}
%
\vspace{-2mm}

However, isobaric models provide us with useful tools, if other
more appropriate formalisms allow us to single out the most relevant
resonances in the reaction mechanism and determine their couplings in order
to significantly reduce the number of free parameters. 
Such an opportunity is offered to
us by a chiral constituent quark approach, as discussed in the next Section. 
Nevertheless, this latter being a non-relativistic 
formalism, can not be applied to the electroproduction processes, other than at
low $Q^2$ kinematic region. So, a possible scenario could be to pin down the
reaction mechanism in the photoproduction using the constituent quark formalism,
then pick up the most relevant resonances and their couplings extracted {\it via}
the quark model and embody them in the Feynman diagrammatic approach to study
the electroproduction reactions. The capability of Feynman diagrammatic technique
to provide the elementary operators and be used as input into the strangeness 
production on nuclei has
already been proven\cite{nuclei}. Finally, the advent of realistic
elementary operators in line with the above procedure implies coupled-channel
treatments\cite{CC1,CC2}.

In the following Sections, we will focus on the very recent 
$\gamma p \rightarrow K^{+} \Lambda$ data\cite{CLAS,Saphir} and study them {\it via} a chiral 
constituent quark approach based on the broken $SU(6)\otimes O(3)$ symmetry.
%
%
\section{Theoretical Frame}
The starting point of the meson photoproduction in the chiral quark model is the low 
energy QCD Lagrangian\cite{MANOHAR}
\begin{eqnarray}\label{eq:Lagrangian}
{\mathcal L}={\bar \psi} \left [ \gamma_{\mu} (i\partial^{\mu}+ V^\mu+\gamma_5
A^\mu)-m\right ] \psi + \dots
\end{eqnarray}
where $\psi$ is the quark field  in the $SU(3)$ symmetry,
$ V^\mu=(\xi^\dagger\partial_\mu\xi+\xi\partial_\mu\xi^\dagger)/2$ 
and 
$A^\mu=i(\xi^\dagger \partial_{\mu} \xi -\xi\partial_{\mu} \xi^\dagger)/2$ 
are the vector and axial currents, respectively, with $\xi=e^{i \Pi f}$. 
$f$ is a decay constant and the field $\Pi$ is a $3\otimes 3$ matrix,
\begin{equation}\label{eq:Pi}
\Pi=\left| \begin{array}{ccc} \frac 1{\sqrt {2}} \pi^\circ+\frac 1{\sqrt{6}}\eta 
& \pi^+ & K^+ \\ \pi^- & -\frac 1{\sqrt {2}}\pi^\circ+\frac 1{\sqrt {6}}\eta & 
K^\circ \\ K^- & \bar {K}^\circ &-\sqrt{\frac 23}\eta \end{array}\right|,
\end{equation}
in which the pseudoscalar mesons, $\pi$, $K$, and $\eta$, are treated
as Goldstone bosons so that the Lagrangian in Eq.~(\ref{eq:Lagrangian}) 
is invariant under the chiral transformation.  
Therefore, there are four components for the photoproduction of
pseudoscalar mesons based on the QCD Lagrangian,
\begin{eqnarray}\label{eq:Mfi}
{\mathcal M}_{fi}&=&\langle N_f| H_{m,e}|N_i \rangle + \nn \\
&&\sum_j\bigg \{ \frac {\langle N_f|H_m |N_j\rangle 
\langle N_j |H_{e}|N_i\rangle }{E_i+\omega-E_j}+ \nn \\
&& \frac {\langle N_f|H_{e}|N_j\rangle \langle N_j|H_m
|N_i\rangle }{E_i-\omega_m-E_j}\bigg \}+{\mathcal M}_T,
\end{eqnarray}
where $N_i(N_f)$ is the initial (final) state of the nucleon, and 
$\omega (\omega_{m})$ represents the energy of incoming (outgoing) 
photons (mesons). 

The pseudovector and electromagnetic couplings at the tree level are given respectively
by the following standard expressions:
\begin{eqnarray}
H_m~&=~&\sum_j \frac 1{f_m} {\bar \psi}_j\gamma_\mu^j\gamma_5^j \psi_j
\partial^{\mu}\phi_m,\label{eq:Hm} \\
H_e~&=~&-\sum_j e_j \gamma^j_\mu A^\mu ({\bf k}, {\bf r}).\label{eq:He}
\end{eqnarray}

The first term in Eq.~(\ref{eq:Mfi}) is a seagull term.
The second and third terms correspond to the {\it s-} and {\it u-}channels,
respectively. 
The last term is the {\it t-}channel contribution and is 
excluded here due to the duality hypothesis\cite{LS-2}.

The contributions from  the {\it s-}channel resonances to the transition 
matrix elements can be written as
\begin{eqnarray}\label{eq:MR}
{\mathcal M}_{N^*}=\frac {2M_{N^*}}{s-M_{N^*}(M_{N^*}-i\Gamma(q))}
e^{-\frac {{k}^2+{q}^2}{6\alpha^2_{ho}}}{\mathcal A}_{N^*},
\end{eqnarray}
with  $k=|\vk|$ and $q=|\vq|$ the momenta of the incoming photon 
and the outgoing meson respectively, $\sqrt {s}$ the total energy of 
the system, $e^{- {({k}^2+{q}^2)}/{6\alpha^2_{ho}}}$ a form factor 
in the harmonic oscillator basis with the parameter $\alpha^2_{ho}$ 
related to the harmonic oscillator strength in the wave-function, 
and $M_{N^*}$ and $\Gamma(q)$ the mass and the total width of 
the resonance, respectively.  The amplitudes ${\mathcal A}_{N^*}$ 
are divided into two parts\cite{zpl97}: the contribution 
from each resonance below 2 GeV, the transition amplitudes of which 
have been translated into the standard CGLN amplitudes in the harmonic 
oscillator basis, and the contributions from the resonances above 2 GeV
treated as degenerate, since little experimental information is available
on those resonances.

The contributions from each resonance 
is determined by introducing\cite{LS-2,LS-1} a new set of 
parameters $C_{{N^*}}$, and the following substitution rule for the 
amplitudes ${\mathcal A}_{{N^*}}$:
\begin{eqnarray}\label{eq:AR}
{\mathcal A}_{N^*} \to C_{N^*} {\mathcal A}_{N^*} ,
\end{eqnarray}
so that 
\begin{eqnarray}\label{MRexp}{\mathcal M}_{N^*}^{exp} = C^2_{N^*}
 {\mathcal M}_{N^*}^{qm} ,
\end{eqnarray}
where ${\mathcal M}_{N^*}^{exp}$ is the experimental value of 
the observable, and ${\mathcal M}_{N^*}^{qm}$ is calculated in the 
quark model\cite{zpl97}. 
The $SU(6)\otimes O(3)$ symmetry predicts
$C_{N^*}$~=~0.0 for ${S_{11}(1650)} $, ${D_{13}(1700)}$, and 
${D_{15}(1675)} $ resonances, and $C_{N^*}$~=~1.0 for other
resonances in Table~II.  
Thus, the coefficients $C_{{N^*}}$ measure the discrepancies between 
the theoretical results and the experimental data and show the extent 
to which the $SU(6)\otimes O(3)$ symmetry is broken in the process 
investigated here.

\vspace{1mm}
%
%
\noindent Table II. {Resonances discussed in Figs. 1 to 4, with their 
assignments in $SU(6)\otimes O(3)$ configurations, masses, 
and widths.} 
\vspace{-5.3mm} 
\begin{table}[h!]\label{tab:Res}
\caption{Resonances included in our study with their 
assignments in $SU(6)\otimes O(3)$ configurations, masses, 
and widths.}
\label{assign}
\begin{center}
\begin{tabular}{|l|lc|cc|cc|}
\hline  
States & & $SU(6)\otimes O(3)$& & Mass & & Width   \\  
 & & & &  (GeV) & &  (GeV)  \\  \hline        
$S_{11}(1535)$&&$N(^2P_M)_{\frac 12^-}$&& && \\[1ex] 
$S_{11}(1650)$&&$N(^4P_M)_{\frac 12^-}$&&1.650&&0.150 \\[1ex]    
$D_{13}(1520)$&&$N(^2P_M)_{\frac 32^-}$&&1.520&&0.130\\[1ex]    
$D_{13}(1700)$&&$N(^4P_M)_{\frac 32^-}$&&1.700&&0.150\\[1ex]
$D_{15}(1675)$&&$N(^4P_M)_{\frac 52^-}$&&1.675&&0.150\\[1ex]
$P_{13}(1720)$&&$N(^2D_S)_{\frac 32^+}$&&1.720&&0.150\\[1ex]    
$F_{15}(1680)$&&$N(^2D_S)_{\frac 52^+}$&&1.680&&0.130\\[1ex]    
$P_{11}(1440)$&&$N(^2S^\prime_S)_{\frac 12^+}$&&1.440&&0.150\\[1ex]    
$P_{11}(1710)$&&$N(^2S_M)_{\frac 12^+}$&&1.710&&0.100\\[1ex]    
$P_{13}(1900)$&&$N(^2D_M)_{\frac 32^+}$&&1.900&&0.500\\[1ex]    
$F_{15}(2000)$&&$N(^2D_M)_{\frac 52^+}$&&2.000&&0.490\\[1ex] 
\hline     
\end{tabular}
\end{center} 
\end{table}
%

One of the main reasons that the $SU(6)\otimes O(3)$ symmetry is
broken is due to the configuration mixings caused by the one-gluon
exchange\cite{IK}. 
Here, the most relevant configuration mixings are those of the
two $S_{11}$ and the two $D_{13}$ states around 1.5 to 1.7 GeV. The 
configuration mixings can be expressed in terms of the mixing angle
between the two $SU(6)\otimes O(3)$ states $|N(^2P_M)>$  and 
$|N(^4P_M)>$, with the total quark spin 1/2 and 3/2.
To show how the coefficients $C_{N^*}$ are related to the mixing angles, 
we express the amplitudes ${\mathcal A}_{N^*}$ in terms of the 
product of the photo and meson transition amplitudes
\begin{eqnarray}\label{eq:MixAR}
{\mathcal A}_{N^*} \propto <N|H_m| N^*><N^*|H_e|N>,
\end{eqnarray}
where $H_m$ and $H_e$ are the meson and photon transition operators,
respectively. For example, for the resonance ${S_{11}(1535)}$ 
Eq.~(\ref{eq:MixAR}) leads to
\begin{eqnarray}\label{eq:MixAS1}
{\mathcal A}_{S_{11}} \propto 
<N|H_m (\cos \theta _{S}
 |N(^2P_M)_{{\frac 12}^-}> - 
\sin \theta _{S}
|N(^4P_M)_{{\frac 12}^-}>) 
\nonumber\\ 
 (\cos \theta _{S} <N(^2P_M)_{{\frac 12}^-}| -
\sin \theta _{S} <N(^4P_M)_{{\frac 12}^-})|H_e|N>.
\end{eqnarray}
%
Then, the configuration mixing coefficients can be related to the
configuration mixing angles 
\begin{eqnarray}
C_{S_{11}(1535)} &=& \cos {\theta _{S}} ( \cos{\theta _{S}} - 
\sin{\theta _{S}}),\label{eq:MixS15} \\
C_{S_{11}(1650)} &=& -\sin {\theta _{S}} (\cos{\theta _{S}} + 
\sin{\theta _{S}}),\label{eq:MixS16} \\
C_{D_{13}(1520)} &=& \cos \theta _{D} (\cos\theta _{D} - 
\sqrt {1/10}
\sin\theta _{D}),\label{eq:MixD15} \\
C_{D_{13} (1700)} &=& \sin \theta _{D} (\sqrt {1/10}\cos\theta _{D} + 
 \sin\theta _{D}).\label{eq:MixD17}
\end{eqnarray}

\section{Results and Discussion}
The above formalism has been used to investigate the recent data on
the differential cross sections\cite{CLAS,Saphir}, as well as 
recoil\cite{CLAS} $\Lambda$ and beam\cite{LEPS} asymmetries.
The adjustable parameters in this approach are the $KYN$ coupling 
constants and one strength ($C_{N^*}$ in Eq.~\ref{eq:AR}) per resonance 
(Table II). Other resonances in Table~I are included in a compact form
and bear no free parameters.

Figures 1 to 4 show the results for three excitation functions at
$\theta_{K}^{CM}$ = 31.79$^\circ$, 56.63$^\circ$, and 123.37$^\circ$ as a 
function of total center-of-mass energy ($W$). 
The choice of the angles is due
to the data released by the CLAS collaboration\cite{CLAS}.

The full model contains the following terms:
\vspace{1mm}

{\it i) Background} ({\bf Bg}): composed of the seagull, nucleon and hyperons Born
terms, as well as contributions from the excited {\it u-}channel 
hyperon resonances; 

{\it ii) High Mass Resonances} ({\bf HMR}): contributions from the excited resonances 
with masses higher than 2 GeV, handled in a compact form as mentioned earlier;

{\it iii) Resonances}: contributions from the excited nucleon resonances (Table II).
\vspace{1mm}

This model is depicted as full curves in the Figures.
Fig.~1 shows the full model and the two set of data
from CLAS\cite{CLAS}, SAPHIR\cite{Saphir} at the same angles. 
Those differential cross  
section data are compatible at the most
backward angle, but show significant discrepencies at other two angles. 
However, the fitting procedure is driven by the
CLAS data, which bear smaller uncertainties. Given the discrepencies
between the two data set, the model reproduces in a reasonable way the
experimental results.

In figures~ 1 to 4, the full model curves are depicted, while in each
figure contributions from individual resonances (Table II) are singled out. 
An account of those contributions is given below.
%
%
\begin{figure}[b!]
\epsfysize=11.cm 
\centerline{\epsfxsize=4.7in\epsfbox{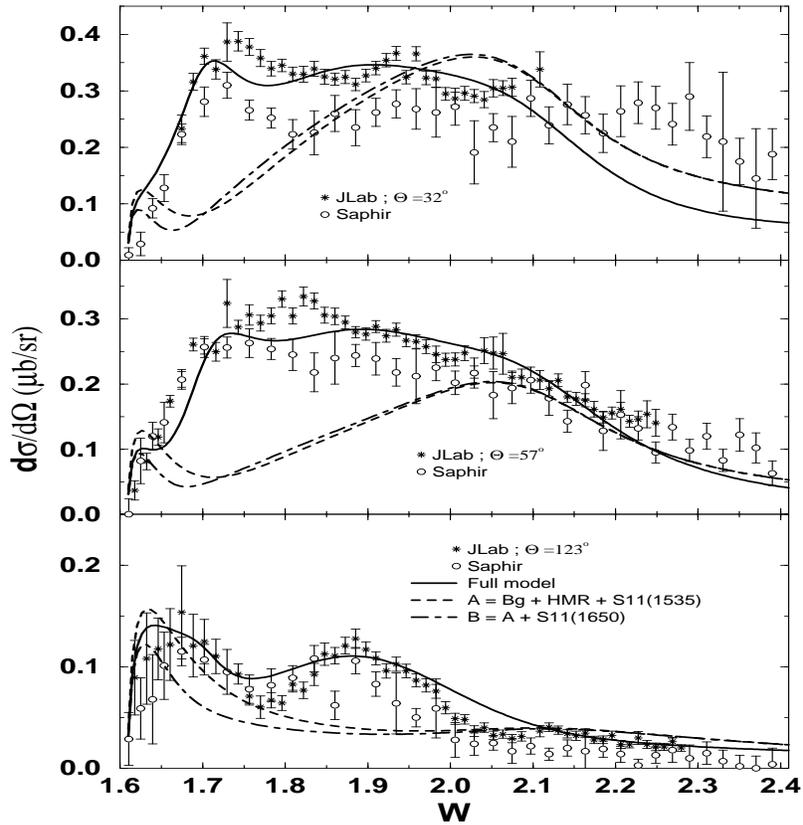}}   
\caption{Differential cross section for the process 
$\gamma p \to K^+ \Lambda$ as a function of total center-of-mass 
energy (W) in GeV. 
The full curves are from the model embodying all known
resonances. Contributions from the background terms, higher mass resonances
plus $S_{11}$(1535) and $S_{11}$(1650) are shown by dashed and 
dash-dotted curves, respectively. 
The  JLab (stars) and SAPHIR (circles) data are
from Refs. [1] and [2], respectively.\label{Fig1}
}
\end{figure}
%

{\bf a) S-wave resonances}

In Fig.1 the dashed curves (A) show the sum of contributions from the background
terms (Bg), High Mass Resonances (HMR), and the $S_{11}$(1535) resonance. A 
peak appears at all three angles close to threshold. The other two terms 
(specially HMR) have large contributions at forward angles and higher
energies. The second
$S_{11}$ resonance, which comes in, due to the configuration mixing, suppresses the
effect of the first resonance and affects very slightly higher energy region
(curves B).

%
\begin{figure}[b!]
\epsfysize=11.cm 
\centerline{\epsfxsize=4.7in\epsfbox{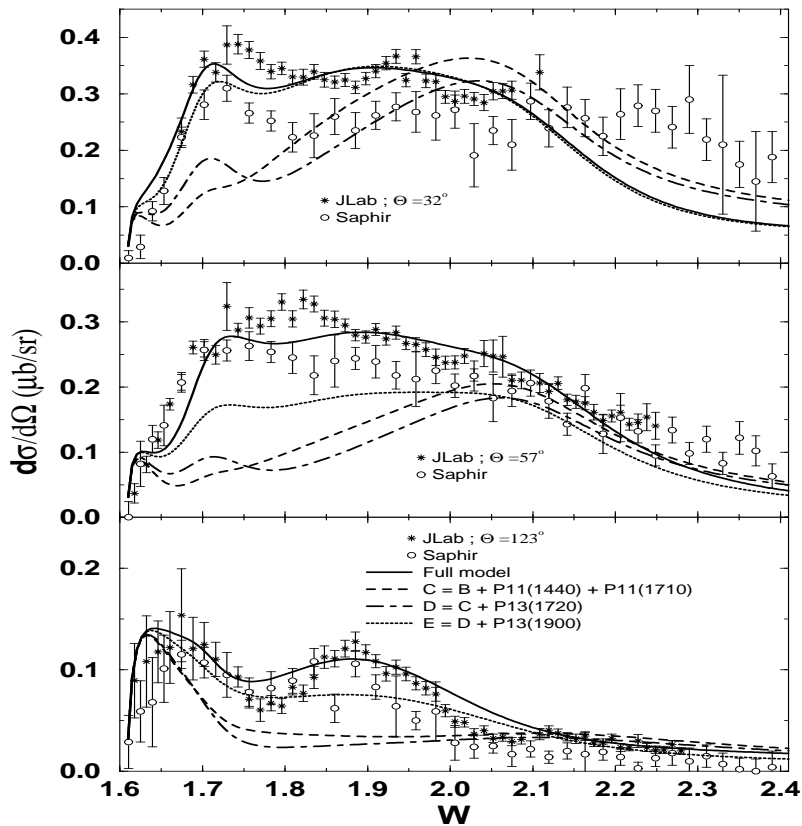}}   
\caption{Same as Fig.~1, but with contributions from P-wave resonances shown.
\label{Fig2}
}
\end{figure}
%
\vspace{2mm}

{\bf b) P-wave resonances}

The Roper resonance, being far below threshold and in spite of its large
width, has no significant contribution. The $P_{11}$(1710) introduces
a tiny structure at forward angles around $W \approx $~1.7 GeV (curves C).
The first $P_{13}$ enhances that structure (curves D). The most dramatic
effect is due to the $P_{13}$(1900). At most forward angle, the curve E gives 
almost the same result as the full calculation, especially above 
$W \approx $1.8 GeV. At the two other angles, roughly half of the cross section
is obtained in the 1.7~$\le W\le$~2.1 GeV region. Below $W \approx $~1.8 GeV,
we witness strong interference phenomena, while the effects around
$W \approx $~1.9 GeV correspond to the (almost) on-shell contributions.
%
%
\begin{figure}[b!]
\epsfysize=11.cm 
\centerline{\epsfxsize=4.7in\epsfbox{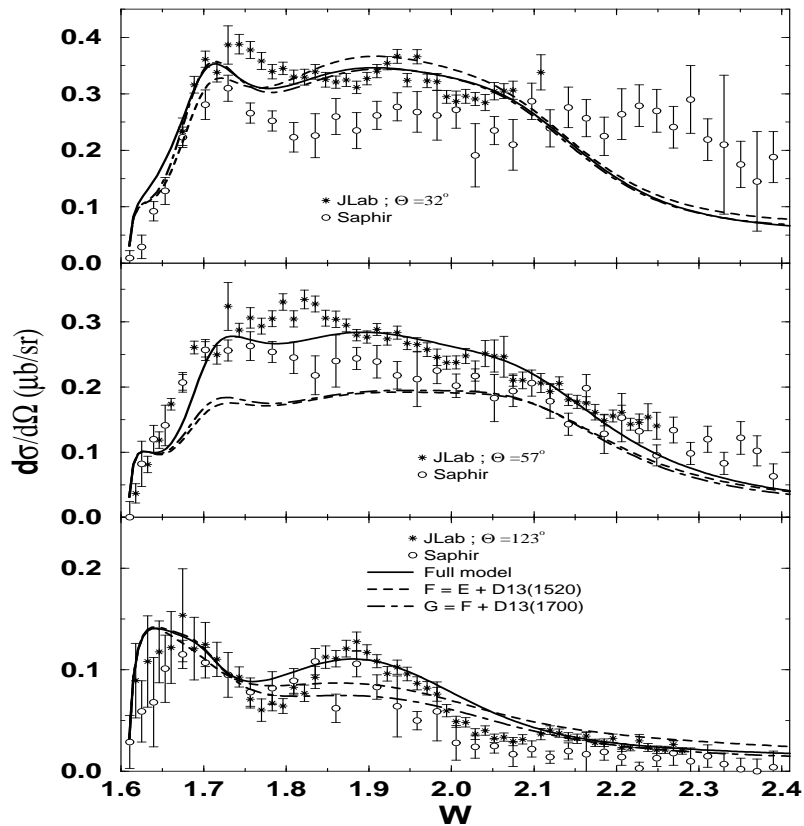}}   
\caption{Same as Fig.~1, but with contributions from $D_{13}$-resonances 
shown.\label{Fig3}
}
\end{figure}
%
\vspace{2mm}

{\bf c) Spin-3/2 D-wave resonances}

The $D_{13}$(1520) and $D_{13}$(1700) affect slightly the extreme angles 
results. The first one (curves F) enhances the cross sections corresponding
to the curves E, while the second one suppresses them with comparable strength.
The final curves G are almost identical to the curves E. Here, $D_{13}$(1700)
contributes again due to the configuration mixing mechanism.

\vspace{2mm} 

{\bf d) Spin-5/2 D- \& F-wave resonances}

The $D_{15}$(1675) shows a noticeable contribution only at the most forward
angle (curves H). In the contrary, the effects of the $F_{15}$(1680) 
appear at two other angles (curves I), and become very important at the 
most backward angle above $W \approx $~1.8 GeV. Here also we are in the 
presence of strong interference mechanisms. Finally, the addition of
the $F_{15}$(2000) leads to the full curves, allowing us to reproduce
data around $W \approx $~1.9 GeV, as well as the high energy part of the data 
at the most backward angle.
%
\begin{figure}[b!]
\epsfysize=11.cm 
\centerline{\epsfxsize=4.7in\epsfbox{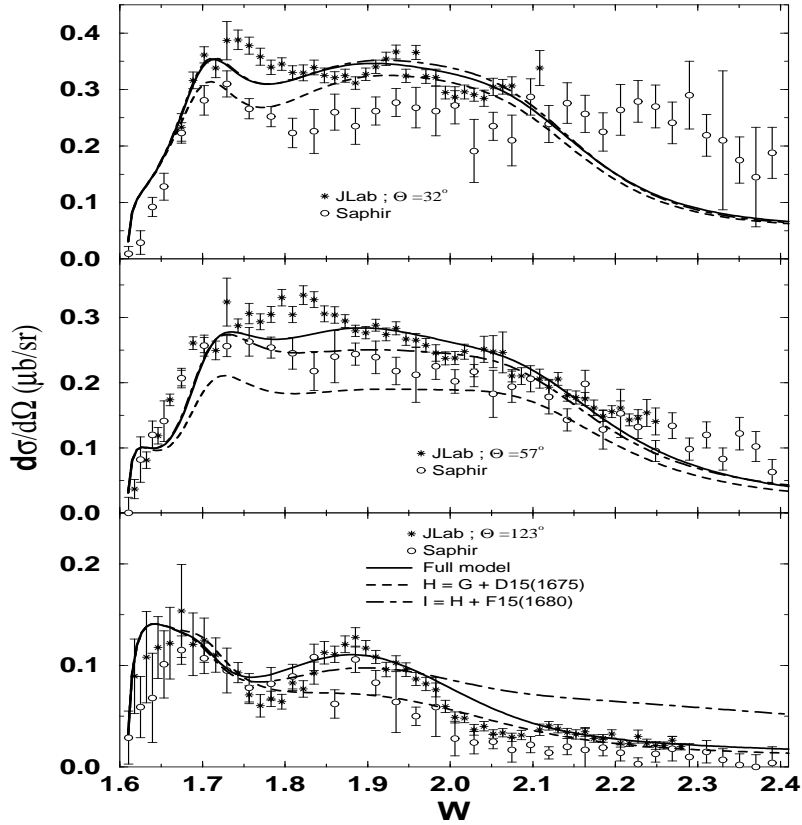}}   
\caption{Same as Fig.~1, but with contributions from $D_{15}$- \&
$F_{15}$-resonances depicted.
\label{Fig4}
}
\end{figure}
%
%
\section{Summary and Concluding remarks}
In this contribution, the preliminary results of a chiral constituent 
quark model have been
compared with the most recent excitation functions measurements at 3 angles from
CLAS\cite{CLAS} and SAPHIR\cite{Saphir}. The discrepencies between the
two data set do not allow to reach strong conclusions on the underlying
reaction mechanism. However, the role played by higher spin and higher mass
resonances, such as $P_{11}$(1900), $F_{15}$(1680), and $F_{15}$(2000) is
established. The obtained model reproduces the single polarization asymmetries
from CLAS and LEPS. Those results, shown during the Conference, could not be
reproduced here because of lack of space.

To go further, several directions deserve attention and effort. From
experimental side, single and double polarization data, e.g. being analyzed
by the GRAAL collaboration, will very likely shed a valuable light on the
reaction mechanism issues. 
From theoretical point of view, the following points need to be 
studied\cite{LSY}:

i) The same formalism should be used to interpret the data on the 
$K^+ \Sigma^\circ$ channel. The forthcoming data from LNS\cite{LNS} on the
$K^\circ$ will also put more constraints on the models.

ii) The effect of the third $S_{11}$ resonance, in line with the $\eta p$
final state investigations\cite{LS-2}, has to be studied for the strangeness 
channels. If this latter resonance has a molecular structure\cite{LW96}, 
it should
show up very clearly in the strangeness production processes.

iii) Given that the SAPHIR data go beyond the resonance region, to explain
highest energy data, one needs very likely to introduce the {\it t-}channel
contributions. For the same reason, explicit investigation of resonances
with spin $\ge$~7/2 might be relevant.

Once such improvements to the quark models are ensured, then the coupled channel
effects\cite{CC2} have to be considered. The couplings extracted within
the coupled-channel formalisms can then be embodied in the isobaric approaches,
including a reasonable number of resonances, to produce the needed elementary operators
and study the electroproduction on both proton and nuclei.

\bigskip

It is a pleasure for me to thank the organizers for their kind invitation 
to this very stimulating conference.
I am indebted to K.H. Glander and R. Schumacher for 
having provided me with the SAPHIR and CLAS data, respectively, 
prior to publication. I am grateful to my collaborators
Z. Li and T. Ye.

\end{document}